\documentclass[runningheads,fleqn]{svmult}

\usepackage{makeidx}   
\usepackage{graphicx}  
\usepackage{subeqnar}  
\usepackage{multicol}  
\usepackage{physproc}  
\makeindex             

\newcommand {\be} {\begin{equation}}
\newcommand {\ee} {\end{equation}}
\newcommand {\Be}{\begin{eqnarray*}}
\newcommand {\Ee} {\end{eqnarray*}}
\newcommand {\bey} {\begin{eqnarray}}
\newcommand {\eey} {\end{eqnarray}}

\newcommand{\comment}[1]{}

\newcommand{\e}{\varepsilon}

\usepackage{amsmath}   
\begin{document}
\title*{Stable chaos}
\toctitle{Stable chaos}
%
%
\titlerunning{Stable chaos}
%
\author{ Antonio Politi\inst{1,2} \and Alessandro Torcini\inst{1,2,3}
}
\authorrunning{A. Politi and A. Torcini}
%
%
\institute{
Istituto dei Sistemi Complessi, Consiglio Nazionale
delle Ricerche, via Madonna del Piano 10, I-50019 Sesto Fiorentino, Italy
\and
Centro Studi Dinamiche Complesse, via Sansone 1, I-50019 Sesto Fiorentino, Italy
\and
Istituto Nazionale Fisica Nucleare - Sezione di Firenze, via Sansone 1, I-50019 Sesto Fiorentino, Italy
}

\maketitle              

\begin{abstract}
{\it Stable chaos} is a generalization of the chaotic behaviour exhibited
by cellular automata to continuous-variable systems and it owes its name to 
an underlying irregular and yet linearly stable dynamics.  In this review we
discuss analogies and differences with the usual deterministic chaos and
introduce several tools for its characterization. Some examples of
transitions from ordered behavior to stable chaos are also analyzed to 
further clarify the underlying dynamical properties. Finally, two models are
specifically discussed: the diatomic hard-point gas chain and a network of
globally coupled neurons.
\end{abstract}

\section{Introduction}

Chaos is associated with an exponential sensitivity of the evolution to tiny
perturbations in the initial conditions, so that the presence of at least one
positive Lyapunov exponent is considered as a necessary and sufficient
condition for the occurrence of irregular dynamics in deterministic dynamical
systems ~\cite{lorenz}. In fact, the first observation in coupled-map models of
stochastic-like behaviour accompanied by a negative maximum Lyapunov exponent
came as a big surprise \cite{CK88,PLOK93}. In order to highlight the unexpected
coexistence of {\it local stability} and {\it chaotic behaviour}, the phenomenon
was called \emph{stable chaos} (SC). Although the definition sounds like an
oxymoron, in practice, there is no logical inconsistency, as the irregular
behaviour is a transient phenomenon that is restricted to finite-time
scales. In spite of this restriction, SC is both a well defined and meaningful
concept, because the transient duration diverges exponentially with the system
size and is therefore infinite in the thermodynamic limit. Moreover, the
stationarity of SC \cite{PLOK93} suggests that it can represent an interesting
platform for studying non-equilibrium phenomena.
A better understanding of SC can be gained by exploring the analogy with the
chaotic behaviour exhibited by elementary cellular automata \cite{bonaccini},
another phenomenon that can be formally defined only in the thermodynamic
limit. In fact, as we clarify in this review, SC is a sort of extension of
cellular-automaton dynamics to continuous-variable systems. In particular, the
spreading velocity of localized perturbations, a standard indicator used to
quantify the degree of chaoticity in cellular automata, proves rather fruitful
also to characterize SC. However, in this latter case it is necessary to
distinguish between finite and infinitesimal perturbations (the latter ones
cannot even be defined in cellular automata, because of the discreteness of the
local variable) and it is thereby possible to define two conceptually different
propagation velocities. This allows giving a fairly general definition of SC as
that of a dynamics dominated by ``finite amplitude" pertubations \cite{PT94,TGP95}.

Altogether, one can express the relevant difference between deterministic
chaos and SC by referring to the relevant flux of information: while in the
former context, information flows from the least towards the most significant
digits, in the latter, it flows from the boundaries towards the core of the
system. It would be therefore desirable to develop a general formalism able to
encompass both phenomena. A promising idea is based on the introduction of
``non democratic" norms which attribute increasingly small weights to the sites
that are increasingly ``far" from the region of interest. Although this
approach allows quantifying the spatial information flow, it can be hardly
extended to account for perturbations that have locally a finite amplitude,
the analysis of which would require a genuine nonlinear treatment. In fact, a
tool like finite amplitude Lyapunov exponents (FALEs) \cite{ABCPV96} appears to be
more appropriate for characterizing SC, although it is not clear how to go
beyond the maximal exponent (for the absence of a proper scalar product
definition in this context).

As we have mentioned above, in systems with a finite number of degrees of
freedom, SC is a transient phenomenon. One might therefore think of using tools
and ideas developed for the characterization of transients such as those 
extensively discussed in the nice review by T\`el and Lai~\cite{tel_review}. One
must however distinguish between SC and standard chaotic transients (for a
seminal paper on the subject, see \cite{grebogi}). In the former case, the
maximum Lyapunov exponent is positive and formulas such as
Kapral-Yorke and Pesin relations can be invoked to express some properties of
the invariant measure in terms of the Lyapunov exponents \cite{ott}. In SC, a
straightforward application of the same formulas yields manifestly useless
predictions, as they do not take into account the spatial information flow that
is the key mechanism of SC. Accordingly, one must still heavily rely on
direct numerical simulations to infer the structure of the invariant measure.
Nevertheless, we suspect that a possible common property of chaotic transients
and SC is the presence of a strange repeller. In fact, chaotic transients are 
almost by definition the manifestation of trajectories evolving in the vicinity 
of a repeller, possibly characterized by a small escape rate \cite{tel_review,ott}. 
This property seems to clash with the absense of unstable orbits in most of the 
models exhibiting SC. However, such models are also characterized by 
discontinuities in phase-space and here below we argue that their smoothing
gives birth to a web of unstable orbits. We are thereby lead to conjecture 
that even though SC is accompanied by a negative Lyapunov exponent, its very
existence requires the presence of topological chaos, i.e. of a finite
topological entropy. When and whether the resulting transient dynamics is
linearly stable or unstable remains however to be clarified.

The review is organized in the following way. In Sec.~\ref{sec:model} we briefly
introduce the reference models that have mostly used to characterize SC. In section
\ref{sec:definition} we properly define SC from the scaling behaviour of the
transient length and discuss its properties in terms of space-time correlations
and fractal dimensions. Then, in Sec.~\ref{sec:ca}, we discuss the relationship
with cellular automata by suitably encoding the space-time pattern. In
particular, we focus our attention on the indeterminacy of the next symbol as a
way to quantify the difference between the original dynamics and that of a
suitable deteministic automaton rule. We also introduce and estimate the
propagation velocity of localized perturbations. In Sec.~\ref{sec:dc}, we
compare SC with the usual deterministic chaos. This is done by smoothing an
otherwise discontinuous coupled-map model and studying the dependence of
standard indicators such as the maximum Lyapunov exponent on the smoothness
of the dynamic rule. As a result, we identify two thresholds: (i) the first one
separates the regions with postive and negative Lyapunov exponent; (ii) the
second, larger, threshold separates the region where finite-amplitude
perturbations propagate faster than infinitesimal ones, from that where the
two velocities coincide with one another (which is the signature of a standard
chaotic evolution). Moreover, we compute the multifractal spectrum of the
Lyapunov exponent, showing that it has a positive tail even when the average
exponent itself is negative. In section \ref{sec:oc} we discuss various
order-to-chaos transitions. In fact, the analogy with cellular automata  
reminds us that such rules are not necessarily chaotic. The intrinsic 
absence of a continuous parameter makes it impossible to investigate 
order-to-chaos transitions in the cellular-automata context. As this restriction 
does not apply to SC, it makes sense and it is desirable to investigate the 
onset of chaotic dynamics in this latter context. We first study a coupled-map 
lattice, the coupling strength being the relevant control parameter. The 
analysis reveals the existence of a fuzzy transition region, where regular and 
irregular dynamics alternate in a complex manner ~\cite{cecconi}. A simple 
stochastic model is then introduced to gain some further insight. 
In the new setup, the transition is of directed-percolation type
\cite{gino}.

For a long time, SC has been found only in abstract mathematical models,
characterized by the presence of discontinuities or nearly discontinuous
\footnote{See the next section for a clarification of this concept.}
evolution rules. This restriction has therefore casted some doubts on the
physical relevance of this phenomenon. In Sec.~\ref{sec:mrm} we discuss a
mechanism that can generically lead to discontinuities in physically
meaningful contexts. The mechanisms requires just the presence of 
$\delta$-like events such elastic collisions between particles or spike 
emissions by neurons. The ``non-commutativity" of such events represents
a genuine source of discontinuities, which may, in turn, give rise to SC. A
diatomic hard-point gas and a network of coupled neurons are discussed in
Sec.~\ref{sec:mrm} as examples of such dynamical systems. The neural network 
model allows us also to discuss an order-to-chaos transition that appears to 
be of ``standard" type (the critical region reduces to a single point), 
although the universality class has not yet been identified~\cite{zillmer1}. 
Finally the still open problems are briefly summarised in 
Sec.~\ref{sec:conclusion}.

\section{Models}
\label{sec:model}

Most of the numerical studies of {\it stable chaos} have been carried out in
a 1D lattice of diffusively coupled maps \cite{cml}, 
\begin{equation}
\label{eq:map1}
x_i(t+1) = (1-\e) f(x_i(t)) +\frac{\varepsilon}{2}[f(x_{i-1}(t)+f(x_{i+1}(t))]
\end{equation}
where $\varepsilon \in [0:1]$ is the coupling constant and the map of the
interval  $f$ is piecewise linear,
\begin{equation}
\label{eq:fun}
f(x) = \begin{cases} 
p_1x + q_1 &  0\le x\le x_c \\ 
1 - (1 - q_2)(x-x_c)/\eta & x_x< x< x_c+\eta \\
q_2 + p_2(x - x_c -\eta)   &  x_c + \eta < x \le 1 ,
\end{cases}
\end{equation}
where $x_c = (1-q_1)/p_1$. The model \cite{PT94} is a continuous
generalization of the systems analysed in \cite{CK88} and \cite{PLOK93}, 
which basically correspond to $\eta=0$, i.e. to a two-branch maps.
The map is continuous because the left and right limits 
in the both connecting points do coincide 
($f(x_c^-)=f(x_c^+)$, $f(x_c^-+\eta)=f(x_c^++\eta)$). Occasionally
in this review we speak of ``quasi-discontinuous" models, implying the
presence of large but localized (in phase-space) amplifying regions.
In this context, this amounts to assuming a small but non-zero width
$\eta$ for the middle branch.  In next the section we restrict 
our analysis to the case $\eta=0$.

Since this model is rather artificial (no specific physical problem lies behind
the choice of $f$, which has been mostly selected for simplicity reasons and
for coherence with the seminal paper \cite{buni}), we find it convenient
to consider a second type of model, namely a chain of Duffing oscillators
\begin{equation}
\label{eq:duf}
\ddot x_i = -\gamma \dot x_i -x_i^3 +D(x_{i-1}+x_{i+1}) + (1+G(t)\sin 2\pi t/T_1)x_i
\end{equation}
where $\gamma$ controls the dissipation, $D$ the diffusion between nearby sites
and $G(t)$ is the modulation amplitude that is periodically switched on and off;
$G=A$ for $\mod(t,T)<T_1$ and zero otherwise.
As discussed in \cite{bonaccini}, for $T_2$ long enough, the Lyapunov exponent is
negative, so that the evolution must eventually converge towards a periodic
orbit, as it indeed does. 

\section{Definition and characterization of stable chaos}
\label{sec:definition}

\begin{figure}[ht]
  \begin{center}
    \includegraphics[clip=true,width= 9cm]{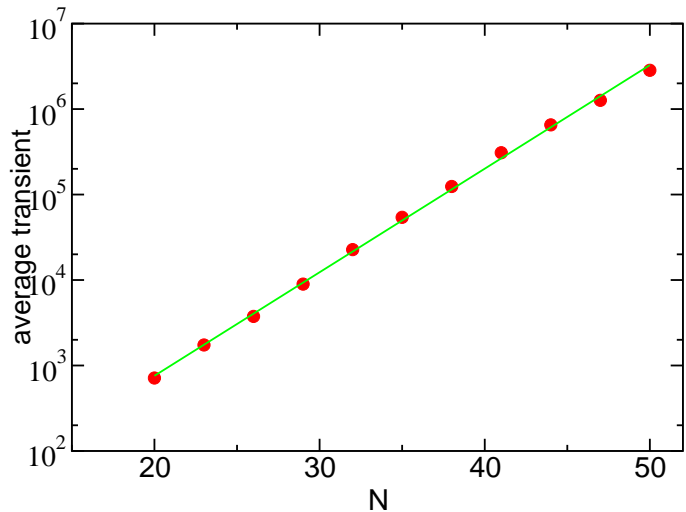}
  \end{center}
\caption{Average transient duration versus the chain length for the diffusively
coupled lattice of maps (\ref{eq:fun}), for $p_1=2.7$, $q_1=0.$, $\eta=0$,
$q_2=0.07$, $p_2=0.1$, and for the coupling strength $\e=2/3$.
}
\label{fig:explo}
\end{figure}

Simulations of the above defined map have revealed the existence of long-lasting
transients followed by a sudden convergence towards some periodic orbit. This
suggests, and simulations confirm, that the basin of attraction of such
orbits is so intricate that the convergence is exponential only for distances 
{\it homogeneously} smaller than some threshold $\theta$. However, since
there exist many different periodic orbits, one cannot estimate the transient
length by determining the distance from an a priori unknown final state.
One can nevertheless determine the distance $d(t,\tau)$ of the configuration
$\{ x_i(t)\}$ ($i=1,N$) at time $t$ from any previous configuration at time
$\tau<t$. As soon as there exists a $\tau$-value such $d(t_c,\tau) < \theta$, we
can conclude that $d(t+t',\tau+ t')$ will tend to 0, indicating that
the dynamics converges towards an orbit of period $t-\tau$. As shown in
Fig.~\ref{fig:explo}, the average (over different choices of the initial
conditions) transient time may increase exponentially with the chain length,
i.e. with the phase space dimension. This indicates that in the thermodynamic
limit, the relevant dynamical regime is not the asymptotic periodic behavior
that is practically unreachable, but what one would naively consider a transient
regime.
This scenario is reminiscent of the disordered regime in directed percolation,
which, in finite systems, has necessarily a finite lifetime, as the dynamics 
sooner or later is absorbed by the homogeneous state \cite{Gras,haye}. 
Irrespective of this difficulty, the disordered regime is a true ``phase"
in the statistical-mechanics sense, as it is stable in infinite systems,
i.e. when the thermodynamic limit is taken before the infinite-time limit.
It should, however, be noticed that in SC the ``absorbing state" is not just a
single homogenous configuration, but may be a set of different and possibly
exponentially long orbits. 

A second striking character of the transient is that the maximum Lyapunov
exponent turns out to be negative. Like for the very existence of SC, this
statement is formally correct only under the assumption of taking first
the thermodynamic limit. In practice, it is sufficient that the transient
duration is long enough to guarantee a good statistical convergence. From the
data reported in Fig.~\ref{fig:explo}, one can see that this is not a limitation
at all, since already in a lattice of 100 maps, the periodic state is
practically unreachable.

\begin{figure}[ht]
  \begin{center}
    \includegraphics[clip=true,width=9cm]{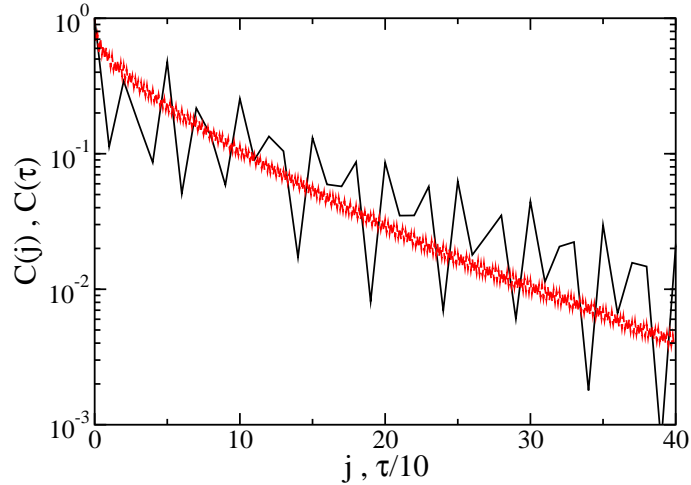}
  \end{center}
\caption{Spatial and temporal (smoother curve) correlations for the same parameter
values of the single map as in Fig.~\ref{fig:explo} and $\e=0.608$.}
\label{fig:corr}
\end{figure}

The very fact that the transient is Lyapunov-stable makes it substantially
different from the chaotic transients that have been often found and attributed
to the existence of some chaotic saddle of high dimensionality \cite{tel_review}.
This is all the way more surprising once we notice that the ``transient"
dynamics is far from regular. In fact, simulations reveal that both spatial and
temporal correlations decay exponentially. An example is reported in
Fig.~\ref{fig:corr}, where we plot
\begin{equation}
C(j) =   \frac{|\langle x_i(t) x_{i+j}(t)\rangle|}{\langle x_i(t)^2\rangle } \qquad , \qquad
C(\tau)= \frac{|\langle x_i(t) x_{i}(t+\tau)\rangle|}{\langle x_i(t)^2\rangle }
\end{equation}
where $\langle \cdot \rangle$ denotes an ensemble average

It is natural to characterize the invariant measure also in terms of its fractal
dimension. Since the whole Lyapunov spectrum is negative, one cannot invoke
the Kaplan-Yorke \cite{ky} formula to predict the number of active degrees of
freedom. Actually, such a formula would imply that the dimension is equal
to zero and this is in fact true for the asymptotic attractor. Therefore, we
must rely only on direct numerical computations.
\begin{figure}[ht]
  \begin{center}
    \includegraphics[clip=true,width=9cm]{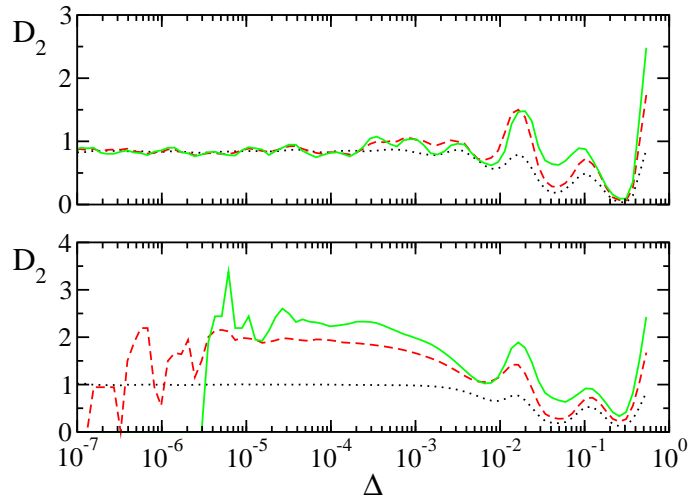}
  \end{center}
\caption{Correlation dimension of the spatial embedding for $\e=0.6008$ (upper
panel) and $\e=0.608$ (lower panel). Dotted, dashed and solid curves correspond
to embedding dimension $e=1$, 2, and 3, respectively.}
\label{fig:fract}
\end{figure}
More precisely, we have decided to compute the correlation dimension \cite{GP}
of spatial sequences of variables \cite{grass89}. In other words, we have
constructed embedding spaces of the type $x_i(t),x_{i+1}(t),\ldots,x_{i+e}(t)$,
for $e=1,2,3$. In each case, we have counted the number of pairs of points
${\mathcal N}(e,\delta)$ that are separated by a distance larger than $\Delta$
in a space of dimension $e$.
Afterwards, we have determined the dimension as the effective derivative, i.e.
\begin{equation}
D_2(e,\Delta)  =\frac{\partial \log {\mathcal N}}{\partial \log \Delta}
\end{equation}
Formally, the correlation dimension is the limit of $D_2(e,\Delta)$ for
$\Delta \to 0$. As for small $\Delta$, $\mathcal N$ is affected by
statistical fluctuations due to the finite number of points, the relevant
question is whether the limiting behaviour sets in for distances that
are numerically accessible. In Fig.~\ref{fig:fract} we report the results
for two different values of the coupling strength, $\e = 0.6008$ and $\e=0.608$,
which correspond to an ordered and chaotic regime, respectively. 
Even in the ordered regime, the fractal dimension is finite, as revealed 
by the plateau, whose height is independent of the embedding dimension. 
The non-zero value of the dimension reflects the disordered spatial structure, 
i.e. the existence of {\it spatial chaos}. Therefore, already from this simple 
case we can conclude on the necessity to go beyond the standard 
Lypapunov-exponent analysis. In the chaotic regime, the effective dimension is 
larger and grows with the embedding dimension $e$ (see lower panel in 
Fig.~\ref{fig:fract}). However, in the absence of theoretical arguments, we
cannot definitely conclude whether the dimension will saturate for 
$e \to \infty$, indicating the existence of a low-dimensional attractor, 
or whether it diverges, suggesting some form extensivity \cite{grass89}. 

\section{Relationship with cellular automata}
\label{sec:ca}
The existence of a stochastic-like dynamics accompanied by an exponential
contraction of infinitesimally close trajectories suggests an analogy with
the so-called chaotic cellular automata (CA) rules \cite{ca}. In fact, in a
finite lattice, any CA rule must eventually produce a periodic orbit, since the
number of distinct states is finite, namely $B^N$, where $B$ is the number of
states of the local variable and $N$ is the number of lattice sites. What makes
a chaotic rule different from an ordered one is precisely the time needed to
cycle through previously visited states: such a time is exponentially long in
chaotic rules \cite{ca}.

\begin{figure}[ht]
  \begin{center}
    \includegraphics[clip=true,width=5cm]{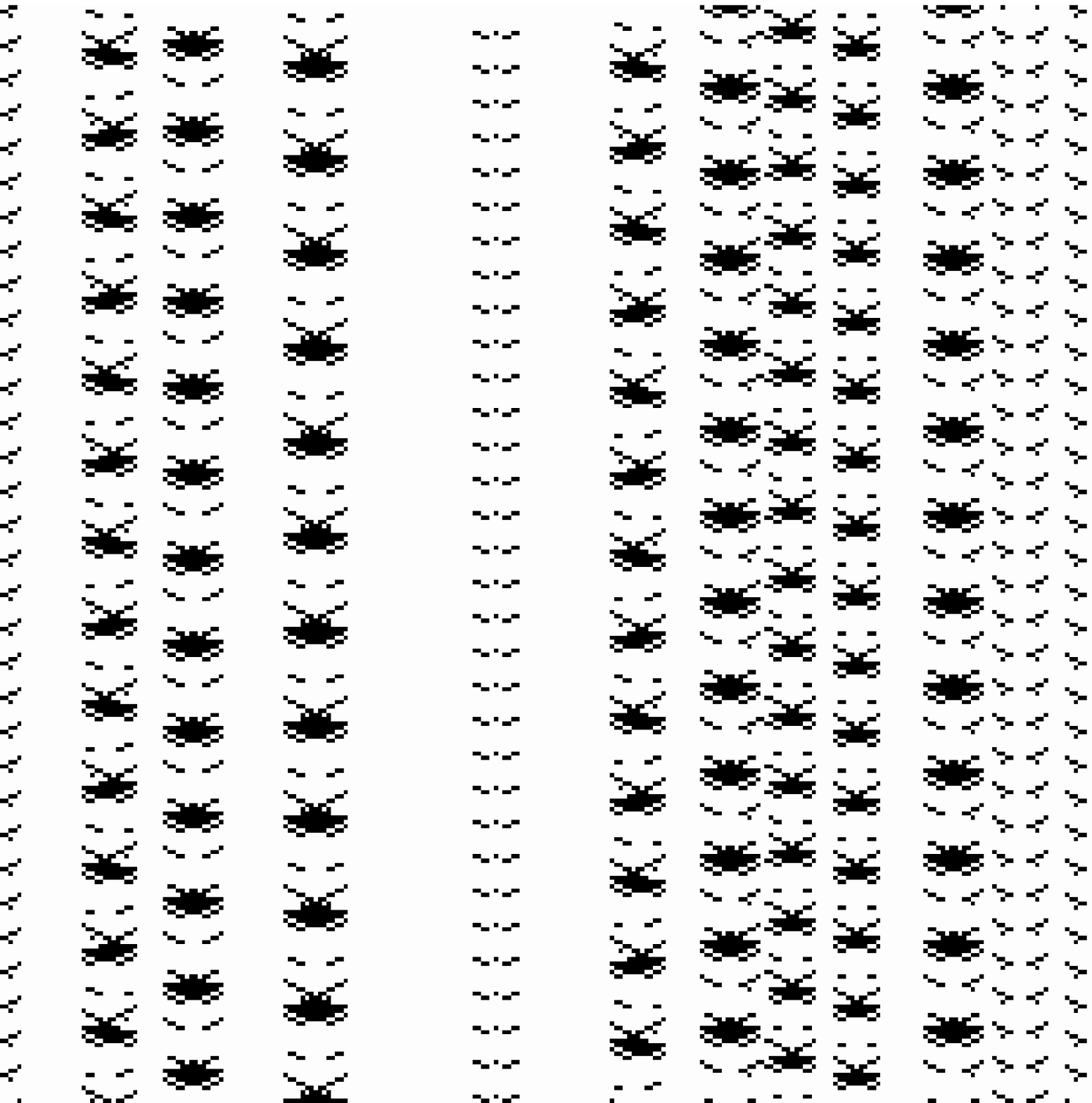}
\hglue .5cm \includegraphics[clip=true,width=5cm]{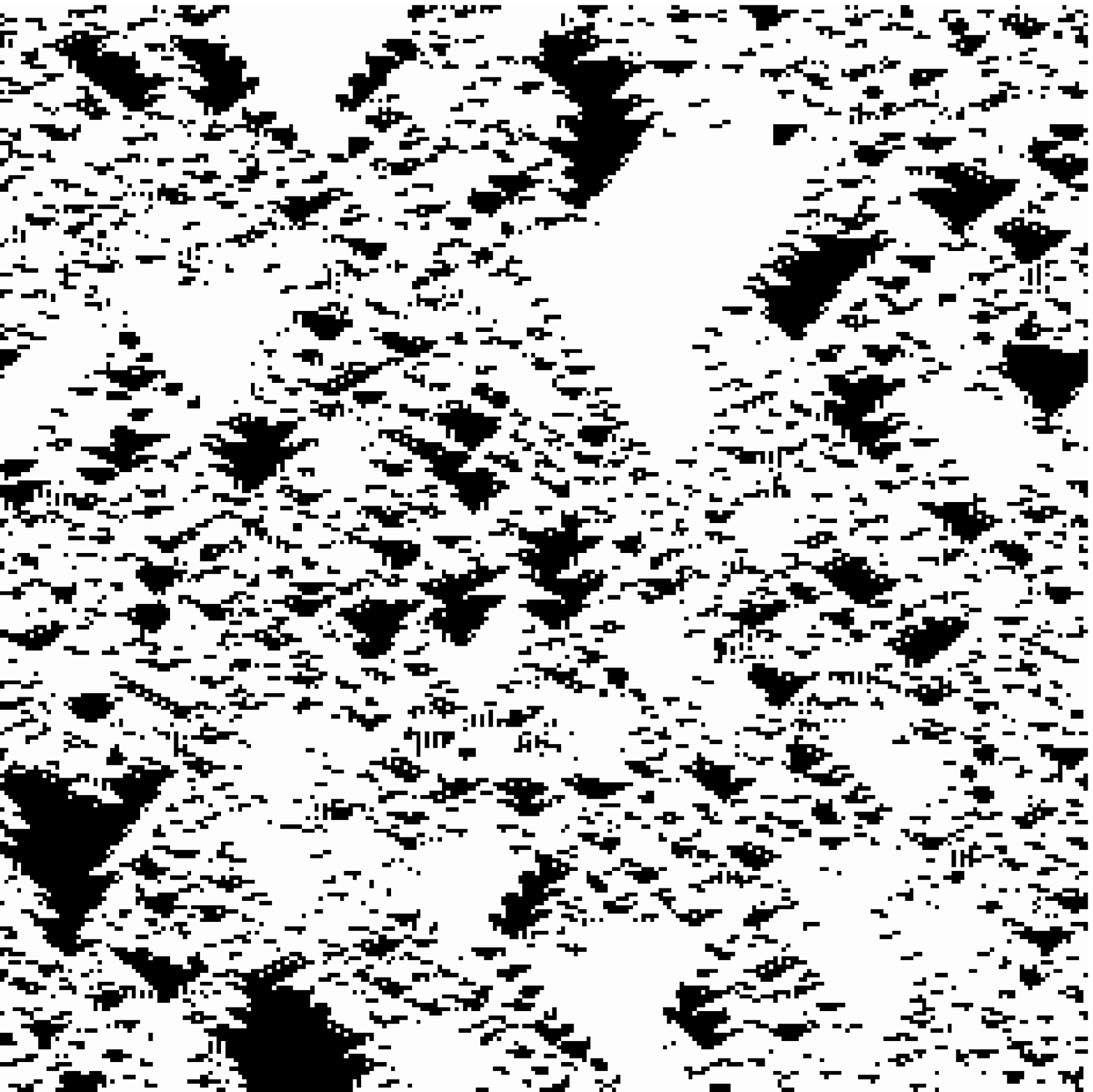}
  \end{center}
\caption{Two patterns generated by iterating Eq.~(\ref{eq:map1}) with the
function $f$ defined as in Eq.~(\ref{eq:fun}) with the same parameter values
as in the previous figures and coupling strength $\e=0.55$ (left panel) and 
0.7 (right panel). Time flows downwards; black corresponds to $x^i(t)<1/2$.}
\label{fig:patt1}
\end{figure}

A binary representation of the dynamics observed in the the coupled map lattice
(\ref{eq:map1}) confirms these expectations. The pattern plotted in
Fig.~\ref{fig:patt1} are indeed very reminiscent of those obtained by iterating
CA rules.

\begin{figure}[ht]
  \begin{center}
    \includegraphics[clip=true,width=9cm]{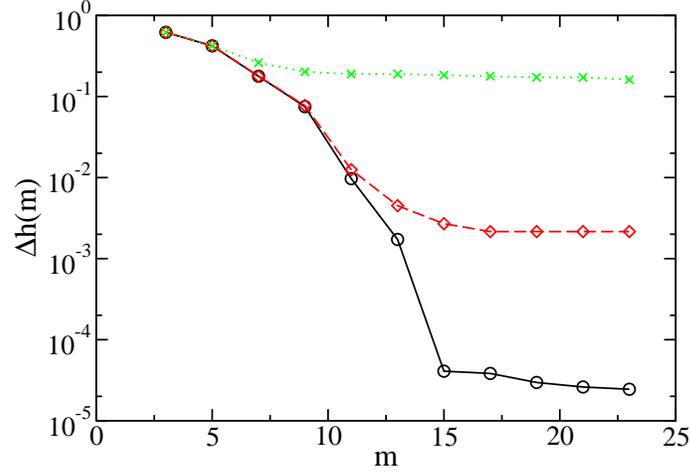}
  \end{center}
\caption{Indeterminacy in the chain of Duffing oscillators (\ref{eq:duf}) with
$\gamma = 0.103$, $D=0.0263$, $\omega = 0.56$, $T_1=T_2/16$ and $T_2=8$, 18, 20
(from top to bottom).}
\label{fig:deltah}
\end{figure}

The relationship with CA can be put on more firm grounds, as we discuss in the
following with reference to the chain of Duffing oscillators (\ref{eq:duf}).
The bistable character of the single oscillators suggests a natural way to
encode the underalying dynamics and thereby to explore possible connections
with CA rules. An appropriate indicator to do so is the indeterminacy
$\Delta h(m)$ of the symbol $s' = s_i(t+1)$,  under the assumption that the
sequence $S(m) =\{s_{i-r}(t),s_{i-r+1}(t),\ldots,s_i(t),\ldots,s_{i+r}(t)\}$,
is observed at time $t$ (time being measured in periods of the forcing term) and
where $m=2r+1$. The indeterminacy is formally defined as \cite{crutch}
\begin{equation}
\Delta h(m) =\sum_S P(S(m)) \sum_{s'}P(s'|S(m)) \log P(s'|S(m))
\end{equation}
where the first sum extends over all sequences of length $m$ generated by the
chain dynamics, and the second sum to the two values of the symbol $s'$.
$P(S(m))$ is the probability to observe anywhere the sequence $S(m)$;
$P(s'|S(m))$ is the conditional probability that the observation of the
symbol $s'$ at time $t+1$ on the site $i$ is preceded by the sequence $S$ at time
$t$ in a window of length $m$ centered around the site $i$.
When the knowledge of $S(m)$ allows to perfectly predict $s'$, then the
indeterminacy is zero. In this case, the symbolic dynamics is perfectly
equivalent to that of a CA defined over a window of length $m$. In
Fig.~\ref{fig:deltah} we report the data for $T_2=8$, 18 and 20. $\Delta h$ is a
non decreasing function of $m$, since the more we assume to know on the past,
the smaller must be the uncertainty on the future. If $\Delta h$ becomes
exactly equal to 0 for a finite $m$, then we can conclude that the dynamics is
perfectly reconstructible from a CA with a finite interaction range. In all
cases we see that for $m$ larger than 15, the curves saturate revealing the
existence of a residual uncertainty. This does neither imply that the 
dynamics contains some degree of stochasticity, nor that the model has to
include longer memory terms. One striking such example was discussed in 
Ref.~\cite{crutch}, where Crutchifield applied this approach to the pattern 
generated by an elementary CA, after it was suitably encoded. The 
indeterminacy of the encoded pattern revealed the presence of a residual 
uncertainty even though the CA rule is deterministic and requires only 
the memory of one past step, while the encoding is even memoryless.  The 
identification of the ``optimal machine" in generic cases is a typical 
example of the hardness of inverse problems. 

\begin{figure}[ht]
  \begin{center}
    \includegraphics[clip=true,width=5cm]{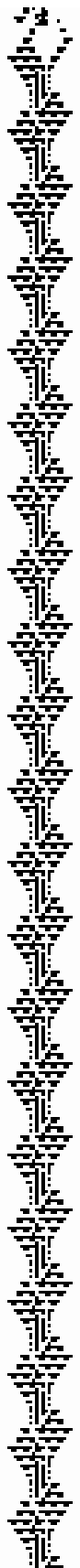}
\hglue .5cm \includegraphics[width=5cm]{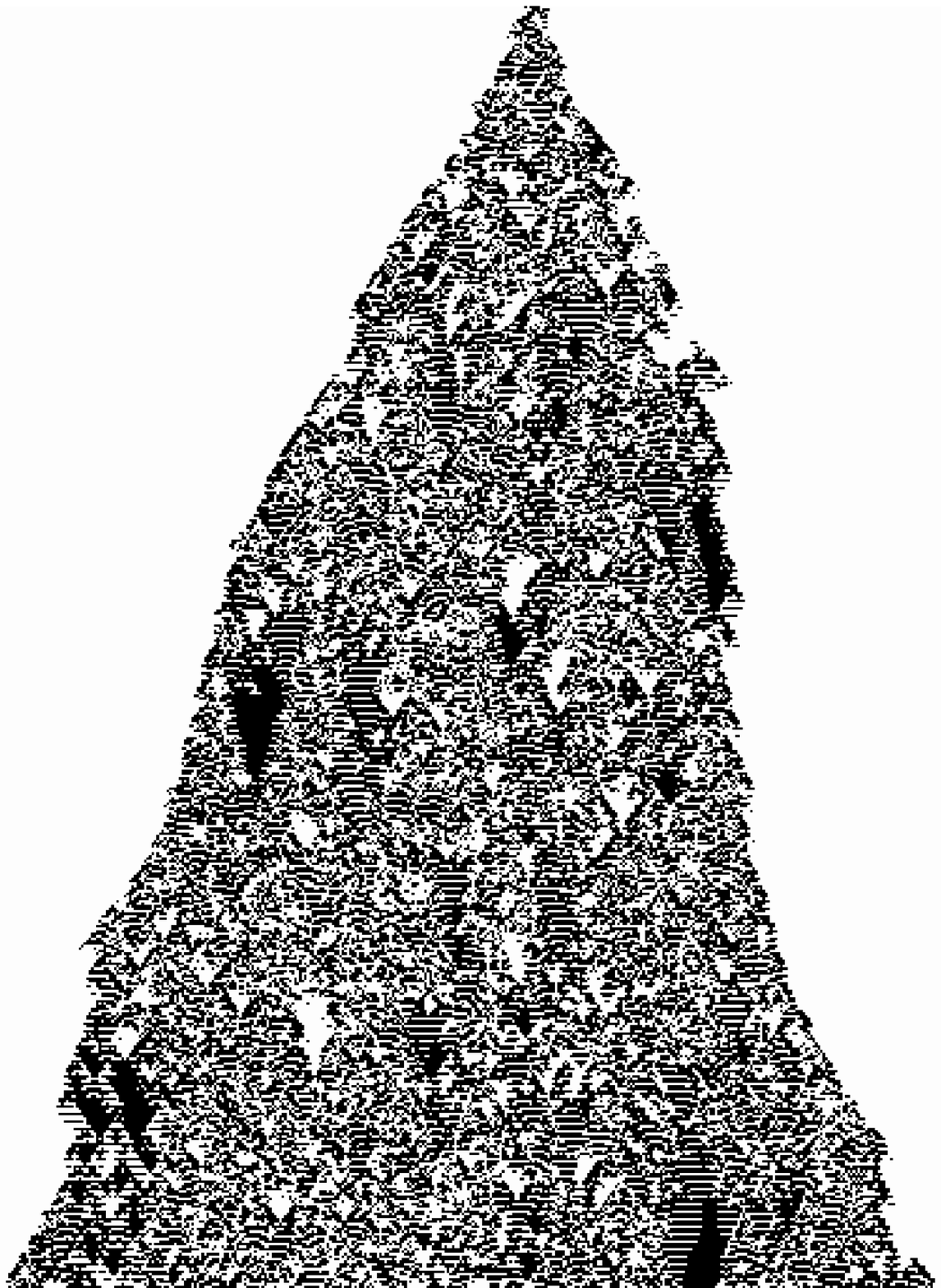}
  \end{center}
\caption{Propagation of initially localized perturbations in the coupled map
lattice for the same values as in Fig.~\ref{fig:patt1}. Time flows downwards.}
\label{fig:diff}
\end{figure}

The analogy with CA suggests to quantify the degree of chaos also in SC  by
determining the velocity $v_F$ of propagation of perturbations. Let us consider
two configurations that initially differ in the interval $[-r,r]$ and let
$i_l(t)$ ($i_r(t)$) denote the leftmost (rightmost) site where they differ more
than some threshold. Accordingly, we can define the front velocity as
\begin{equation}
\label{eq:velonl}
v_{F} = \lim_{t \to \infty} \frac{i_r - i_l}{2t}  .
\end{equation}
Within CAs, a finite spreading velocity is considered as an evidence of
chaotic behaviour \cite{ca}. In fact, this is true also in the context of SC, as
it can be seen in Fig.~\ref{fig:diff}, where we plot the spreading of an initial
difference for parameter values that correspond to ordered and irregular 
behaviour. There we see that the perturbation spreads only in the latter case
(see the right panel).
In the CA language, the velocity $v_f$ is often named the ``Lyapunov exponent"
\cite{ca}. In fact the evolution equation of an elementary CA
can be formally written as a mapping of ${\mathcal R}^2$ into itself,
\begin{equation}
 u^l(t+1) = F^l(u^l(t),u^r(t)) \qquad  u^r(t+1) = F^r(u^l(t),u^r(t)) 
\end{equation}
where $u^r(t) = \sum_{\ge 0} s_i(t) 2^{-i}$, $u^l(t) = \sum_{i<0} s_i(t) 2^{i}$
and $s_i(t) = 0,1$ (for the sake of simplicity we refer to binary automata).
In general $F^l$ and $F^r$ are highly singular functions, but this does not
forbids to define a sort of Lyapunov exponent from the
growth rate of an arbitrarily small perturbation $\delta$. 
In practice, if two configurations differ only in the interval $[-r,r]$,
we can equivalently say that the representation points in ${\mathcal R}^2$
are separated by a distance $\delta \approx 2^{-r}$. Moreover, 
if the spatial region where the two configurations differ increases with a
velocity $v_f$, we can state that the ${\mathcal R}^2$
distance grows as $\exp[ (v_f \log 2)t]$, thus confirming that, apart from
a multiplicative factor, the velocity plays the role of a Lyapunov exponent.
In coupled map lattices, the local variable is continuous rather than binary,
but this does not change the substance of the argument.

\section{Relationship with deterministic chaos}
\label{sec:dc}

The original model where SC has been observed for the first time has a
peculiarity, namely, the discontinuity of the mapping \cite{PLOK93}. As a
result, the distance between two arbirarily close trajectories can suddenly
become of order ${\mathcal O}(1)$, when they find themselves on opposite sides
of the discontinuity. It is therefore reasonable to study the continuous
version of the model, i.e. to assume $\eta \ne 0$ in (\ref{eq:fun}). In the limit
$\eta \to 0$, the map (\ref{eq:fun}) reduces to the original discontinuous system. 

Already at the level of the single map (i.e. without invoking any
spatial coupling), the introduction of an additional branch may drastically
modify the structure of the corresponding dynamical system. This is clear in 
the simple case $q_1=q_2=0$, $p_1=p_2=2$. For $\eta=0$, the topological entropy
is $H = \ln 2$, as the map corresponds the Bernoulli shift; however, for any
arbitrarily small, but finite, $\eta$-value, the appearance of a third branch
induces a jump to $H = \ln 3$. For the parameter values that correspond to the
SC regime discussed in the previous section, the consequence of a finite
$\eta$-value is even more striking, as $H$ is strictly equal to zero for
$\eta=0$, while it is finite for $\eta=0^+$. This can be understood, by
performing a slightly nonconventional symbolic analysis. Let us start by
recalling that for the original parameter values, there exists a stable
period-3 orbit, whose points are ordered as $p_1<p_2<x_1<x_2<p<3<1$. Because of
the third contracting branch, the interval $[x_2,1]$ is asymptotically squeezed
to a point, so that we can identify the leftmost point $x_2$, with the right
border of asymptotically distinct trajectories. Analogously, the interval
$[0,p_1]$ is also squeezed to zero and we can accordingly interpret $p_1$ as the
right border of the relevant interval. Finally $[p_2,x_1]$ is also squeezed to
zero and can be neglected as well. As a result, the relevant dynamics is
described by the mapping of $I_1=[p_1,p_2]$ and $I_2=[x_1,x_2]$: $f(I_1)=I_2$ and
$f(I_2) = I_1\cup I_2$. It is easy to show that the corresponding topological
entropy is the golden mean $H=\log[(1+\sqrt{5})/2]$. Therefore, we can at least
conclude that the introduction of a finite but arbitrarily small $\eta$ induces
topological chaos in an otherwise stable environment.

In the single map, the existence of a fractal chaotic repellor can
induce long transients only for those trajectories that are carefully selected
in the vicinity of the repellor itself. It is reasonable to conjecture that
in spatially extended systems, the stable manifold of the repellor forces
generic trajectories to follow an intricate arrangement of ``channels" before
landing on some periodic orbits. What are, however, the dynamical properties of
the lattice, when finite $\eta$-values are assumed?

\begin{figure}[ht]
  \begin{center}
    \includegraphics[clip=true,width=9cm]{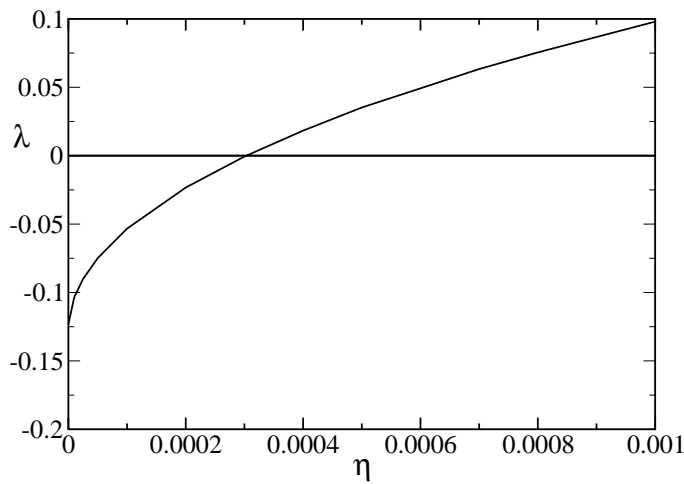}
  \end{center}
\caption{
Maximum Lyapunov exponent of the map lattice (\ref{eq:fun}) as a function of
$\eta$, while the other parameter values are the same as in
Fig.~\ref{fig:explo}. The results have been obtained for $N=200$, but are
practically independent of the system size.}
\label{fig:lyaptot}
\end{figure}

\begin{figure}[ht]
  \begin{center}
    \includegraphics[clip=true,width=9cm]{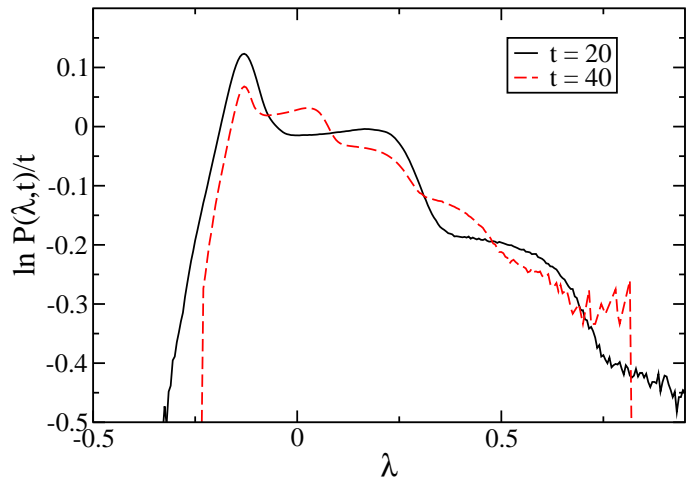}
  \end{center}
\caption{Multifractal distribution of Lyapunov exponents for $\eta=1. 10^{-4}$,
where the average Lyapunov exponent is still negative.}
\label{fig:histo}
\end{figure}

First of all, it important to notice that while the topological enropy jumps
abruptely to a finite value, the Lyapunov exponent exhibits a smooth behaviour
(see Fig.~\ref{fig:lyaptot}), i.e. for sufficiently small $\eta$-values
($\eta <\eta^* < 3.10^{-4}$) it stays negative. This implies that the phenomenon
of SC is {\it generic} (in the mathematical sense), even though the window of 
existence is (at least in this context) rather narrow. More accurate information 
can be extracted by performing a multifractal analysis of the Lyapunov
exponent \cite{multifractal,review_cencini}. In particular, we have computed the
probability  distribution $P(\lambda,t)$ of the maximum Lyapunov exponent
$\lambda$ over a time span $t$. For sufficiently large $t$, the probability
$P(\lambda,t)$ is expected to scale as $P(\lambda,t) =\exp[-G(\lambda)t]$
where $G$ is a dynamical invariant whose operative definition is obtained
by inverting this scaling relation, $G = -(\log P)/t$.
In Fig.~\ref{fig:histo} we have plotted the results obtained for $t=20$ and $40$. 
Even though the most probable and average Lyapunov exponent is negative
(the spectra refer to $\eta = 10^{-4} < \eta^*$), there is a positive tail,
in agreement with the conjectured existence of a web of unstable orbits.
The smoothed steps on the right of the maximum correspond to the number of
times a sample trajectory is actually visiting the expanding branch.
The two spectra do reasonably overlap, suggesting that the time $t=40$ is
already in the scaling regime, although finite-size corrections are still large
(notice, in fat, that the maximum of $G$ has to be, by definition, equal to
0).

Altogether, SC appears to be somehow complementary to the blow-out phenomenon
discovered in the study of synchronization transitions~\cite{PRK02}.
While analysing the stability of the synchronization manifold, it has been
discovered that in some circumstances, the corresponding (multifractal)
Lyapunov spectrum altough mostly confined to the negative semi-axis, may
exhibit a positive tail. In such a case, one has to go beyond the
linear stability analysis, because whenever the distance is amplified,
nonlinear terms are responsible for either bringing the trajectory back towards
the manifold or letting it escape away. In the context of SC, nonlinear terms
bring the trajectory back towards the ``invariant manifold", although the
mechanism is perfectly efficient only in the infinite dimensional limit.

In order to clarify the mechanisms by which nonlinearities contribute to
stabilizing the chaotic dynamics, it is convenient to analyse the propagation
of perturbations. We start by briefly recalling the concept of convective
Lyapunov exponents \cite{DK87}. Given a unidimensional lattice model in the
stationary regime, let us introduce a $\delta$-like perturbation at time $t=0$
in the origin $i=0$ and imagine to monitor the perturbation amplitude $w_i(t)$.
Kaneko and Deissler \cite{DK87} suggested that
\begin{equation}
\label{eq:DK}
w_i(t) = \exp [ \Lambda(i/t) t]
\end{equation}
where $\Lambda(v)$ represents the (exponential) growth rate of a perturbation
in a frame moving with the velocity $v$. It is a priori obvious that for $v=0$,
one recovers the usual Lyapunov exponent, and that for large velocities one has
to expect negative growth rates. In fact, $\Lambda(v)$ has a typically parabolic
shape with the maximum in zero. All velocities for which $\Lambda(v)>0$
correspond to growing perturbations. The limit velocity for linearly
propagating perturbations is fixed by the marginal stability criterion
$\Lambda(v_L)=0$. Instead of determining directly $\Lambda(v)$, it is more
convenient to exploit the chronotopic approach set in \cite{TP92,lepri}, and
formally introduce a perturbation with a spatial amplification factor
\begin{equation}
\label{eq:CT1}
w_i(t) = \hbox{e}^{-\mu i} u_i(t)
\end{equation}
In our lattice model, the evolution rule for $u_i(t)$ reads
\bey
\label{eq:tang1}
u_i(t+1,\mu) &=& \frac{\e}{2}\hbox{e}^{-\mu i} f'(x_{i-1}(t)) u_{i-1}(t,\mu) + \\
& &(1-\e) f'(x_i(t)) u_i(t,\mu) + \frac{\e}{2} \hbox{e}^{\mu i}
f'(x_{i+1}(t))u_{i+1}(t,\mu)
\nonumber
\eey
By iterating this recursive equation with suitable boundary conditions
(periodic conditions are typically optimal, as they reduce finite-size
effects), we obtain the chronotopic growth rate $\lambda(\mu)$. Altogether, an
infinitesimal perturbation $w_i(t) = \exp [\lambda(\mu)t -\mu i]$ with a
spatial growth rate $\mu$ grows in time with an exponent $\lambda(\mu)$.
The evolution of the initially localized perturbation is connected to
$\lambda(\mu)$ by a Legendre transform 
\begin{equation}
\label{eq:legendre}
\Lambda(v) = \lambda(\mu) -\mu \lambda'(\mu) \quad ; \quad v = \lambda'
\end{equation}
In order to determine the velocity corresponding to a given $\mu$-value, it
is necessary to compute the derivative of $\lambda(\mu)$. Since the numerical
computation of derivates is always affected by large numerical errors, it is
convenient to perform a few more analytical steps \cite{TP92}. By introducing,
\begin{equation}
\label{eq:leg-der}
u_i(t,\mu+d\mu) = u_i(t,i) + z_i(t,\mu)d \mu
\end{equation}
in the recursive relation (\ref{eq:tang1}), we obtain an equation for the
deviation $z_i(t,\mu$),
\bey
\label{eq:tangz}
z_i(t+1,&\mu&) = \frac{\e}{2}\hbox{e}^{-\mu i} f'(x_{i-1}(t))
(z_{i-1}(t,\mu)-u_{i-1}(t,\mu)) \\
&+& (1-\e) f'(x_i(t)) u_i(t,\mu) +
\frac{\e}{2} \hbox{e}^{\mu i} f'(x_{i+1}(t)) (z_{i+1}(t,\mu)+u_{i+1}(t,\mu)) \, .
\nonumber
\eey
The knowledge of $z_i$ and of $u_i$ allows determining $\lambda'$. In fact,
by taking the $\mu$ derivative in the definition of the chronotopic Lyapunov
exponent,
\begin{equation}
\label{eq:chrono1}
\lambda(\mu) = \frac{1}{2} \lim_{t\to\infty}  \frac{|| {\bf u}(t)||^2}{t}
\end{equation}
one obtains
\begin{equation}
\label{eq:chrono2}
\lambda'(\mu) = \lim_{t \to \infty} \frac{ {\bf u}(t)\cdot {\bf z}(t)}
{t || {\bf u}(t)||^2}
\end{equation}
where $\cdot$ stands for the scalar product. In order to better understand the
selection process of the propagation velocity, it is convenient to go back to
the evolution of a single exponential profile
$w_i(t) = \exp [\lambda(\mu)t -\mu i]$. Its velocity is obviously
$V(\mu) = \lambda/\mu$. From the Legendre transform we have that
\begin{equation}
\label{eq:velo2}
\frac{dV}{d\mu}= \frac{1}{\mu} \left( \frac{d\lambda}{d\mu} -
\frac{\lambda}{\mu} \right) = -\frac{\Lambda}{\mu^2}  \, . 
\end{equation}
Since the perturbation velocity is identified by the equation $\Lambda=0$, we
see also that it corresponds to the minimum of $V(\mu_0)$. In other words, as
long as the evolution is controlled by linear mechanisms, the slowest among all
possible fronts is selected.
\begin{figure}[ht]
\begin{center}
\includegraphics[clip=true,width=9cm]{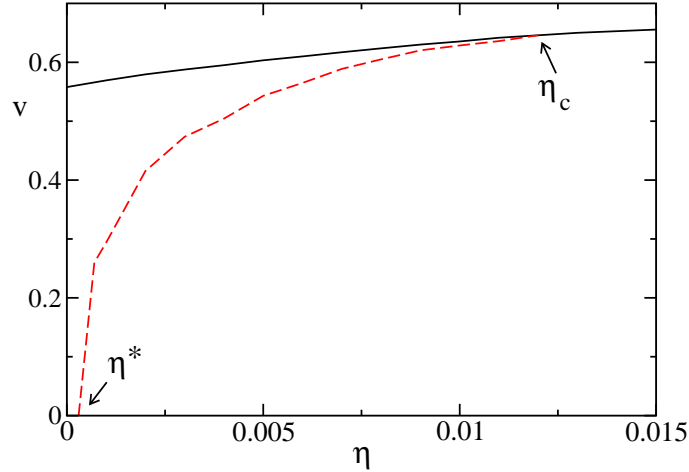}
\end{center}
\caption{Linear ($v_L$, solid curve) and front ($v_F$, dashed curve) velocity 
versus $\eta$. Deterministic chaos exists only for $\eta>\eta^*$. Beyond $\eta_c$, 
$v_L=v_F$.}
\label{fig:velo}
\end{figure}
Let us now turn our attention to fronts delimiting finite perturbations. Since
even such fronts must have a leading infinitesimal edge, $v_F$ will be
$v_F(\mu^*)$ for some $\mu^*$. It is hard to imagine that $\mu^*$ is smaller than
$\mu^0$: accordingly, either $v_F=v_L$ or $v_F >v_L$. This scenario is perfectly
confirmed by the study of the model (\ref{eq:fun}). Solid and dashed curves in
Fig.~\ref{fig:velo} correspond to $v_F$ and $V_L$, respectively. There we see
that $v_F$ is strictly larger than $v_L$ for $\eta<\eta_c \approx 1.2 10^{-3}$,
while above $\eta_c$ the two coincide within numerical accuracy. One can
also notice that the linear velocity is not defined for $\eta< \eta^*$, where
the system is linearly stable and no propagation of infinitesimal perturbations
can occur.

As discussed in Ref.~\cite{{TGP95}}, the mechanism responsible for the finite
difference between $v_L$ and $v_F$ is that perturbations of increasing amplitude
(starting from infinitesimal ones) tend to propagate faster and thereby to push
the corresponding front. These results are indeed fairly general and not just
restricted to the model considered in this section \cite{cencini_torcini_pre}.
Moreover, this phenomenology is conceptually equivalent to that observed in the
context of front propagation (see, e.g. fronts connecting steady states in
reaction-diffusion systems \cite{wim1,wim2}), that is effectively described by
the famous Fisher-Kolmogorov-Petrovsky-Piskunov equation \cite{KPP}.

Altogether we can conclude by stating that the front velocity proves to be a
useful indicator to identify the presence of SC (in spatially extended systems) 
from the presence of nonlinear propagation mechanisms that cannot be accounted
for by linear stability analysis \cite{TGP95}. In such a sense, the results in
Fig.~\ref{fig:velo} indicate that SC persists up to the second threshold and not
just to the first one \cite{PT94}.

\section{From order to chaos}
\label{sec:oc}

Once ascertained that SC is a sort of extension of CA chaos to systems
characterized by continuous variables, it is natural to investigate the
possible phase transitions, a question that cannot even be posed in
CAs, where all variables are discrete. The front propagation velocity
$v_F$ provides the right tool to assess the relative stability of the two
phases. Let us, in fact, consider two initial conditions: a reference
trajectory $\{x_i^{0}\}$, and a perturbed one $\{x_i\}$ differing only in a
finite interval $-L<i<L$, where it is randomly set. If the interval where 
$v(i)$ is of order ${\mathcal O}(1)$ increases by eating the region where 
the field was initially equal to zero, we can conclude that the chaotic 
phase is thermodynamically stable. 

\begin{figure}[ht]
  \begin{center}
    \includegraphics[clip=true,width=10cm]{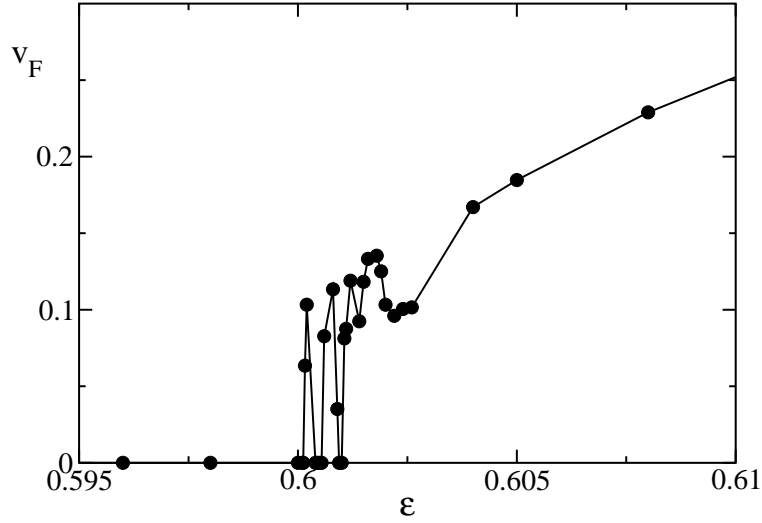}
  \end{center}
\caption{Front propagation velocity in the coupled map lattice
(\ref{eq:map1},\ref{eq:fun}) for the same parameter values as in
Fig.~\ref{fig:explo}.}
\label{fig:velo02}
\end{figure}

In Fig.~\ref{fig:velo02} we plot the results of careful computations
performed with the coupled map lattice (\ref{eq:map1}) for different values of
the coupling strength $\e$. The $\e$-range has been selected so as to include
both the ordered and the chaotic phase. In fact, we see that the front 
velocity is equal to zero (finite) in the left (right) part of the figure.
However, these two clearly distinct phases are not separated by a point-like
transition. We find instead a fuzzy region, where chaos and order alternate in
a seemingly irregular manner. Is this an evidence of the complexity that is
sometimes invoked to exist at the edge of chaos? Pure numerics alone is not
sufficient to provide a convincing answer to such a difficult question.

\begin{figure}[ht]
  \begin{center}
    \includegraphics[clip=true,width=9cm]{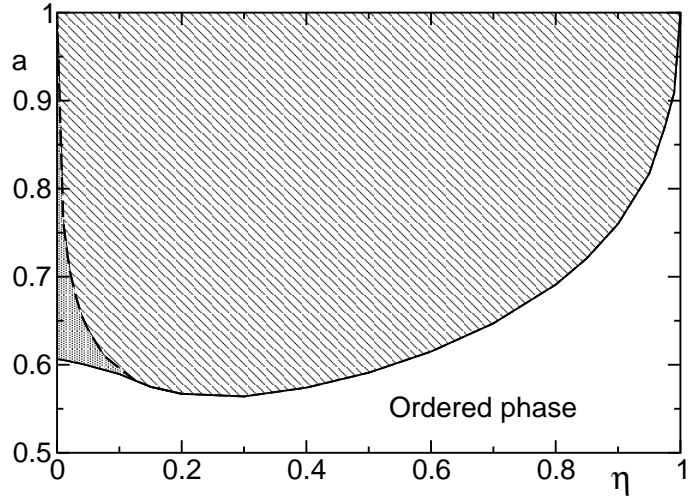}
  \end{center}
\caption{Phase diagram of the stochastic model Eq.~(\ref{eq:ran2b}).}
\label{fig:phase}
\end{figure}

An exact formulation and solution of this problem requires to control
simultaneously two trajectories, a task that is nearly impossible. A simpler
formulation which can nevertheless help to gain some insight on the transition
consists in assuming a random evolution for the reference trajectory, and thus
reducing the problem to that of characterizing the stochastic evolution of
the difference field $v_i(t)$ \cite{gino}. In mathematical terms, this amounts to
studying the equation,
\begin{equation}
\label{eq:ran1}
v_i(t+1) = (1-\e) w_i(t+1)+\frac{\varepsilon}{2}[w_{i-1}(t+1)+w_{i+1}(t+1))]
\end{equation}
where
\begin{equation}
\label{eq:ran2a}
w_i(t+1) = \begin{cases} 
v_i(t)/\eta  & \hbox{w.p.} \qquad p = a \eta \\ 
a v_i(t) & \hbox{w.p.} \qquad 1 - p
\end{cases} \qquad \hbox{if} \quad v_i(t) < \eta
\end{equation}
\begin{equation}
\label{eq:ran2b}
w_i(t+1) = \begin{cases} 
1  & \hbox{w.p.} \qquad p = a v_i(t)\\ 
a v_i(t) & \hbox{w.p.} \qquad 1 - p
\end{cases} \qquad \hbox{if} \quad v_i(t) \ge \eta
\end{equation}
The stochastic $1/\eta$ amplification simulates the effect of visiting the
expanding interval of the map (\ref{eq:fun}). The amplification saturates to
take into account the boundedness of the dynamics. This is the only element
breaking the linearity of $v_i(t)$ dynamics. Moreover, for the sake of
simplicity, we assume that a uniform contraction rate $a$ (an assumption that
is basically equivalent to set $p_1=p_2<1$). At variance with the
original deterministic model, here a detailed numerical analysis of the
parameter space $(a,\eta)$, reveals that ordered and chaotic phases are
separated by a standard phase transition (see Fig.~\ref{fig:phase}) that
belongs to the directed percolation (DP) type for small enough values of $\eta$ 
\cite{GLPT03} and seems to be of multiplicative noise type beyond some critical
$\eta$ value \cite{munoz_review}.

Notice that this stochastic model is even more closely related to the problem
of synchronization between mutually coupled map lattices (see
Refs.~\cite{baroni,ahlers,bagno1,ginelli_2,bagno2,ctt08} for a more 
detailed discussion), since
the assumption of a stochastic evolution is appropriate everywhere in parameter
space including the critical region separating the two phases.

In the SC context, the DP transition is the most relevant one, as it occurs
precisely in the regime where the evolution is characterized by a negative
Lyapunov exponent. DP was introduced and is usually discussed in systems where
the local variable has just two states: 0, and 1. Moreover, the dynamical rule
is such that 1's cannot spontaneously appear in a sea of 0's.
This is the key difference with respect to the present context, where the
variable $v_i$ is continuous and thereby the 0-state is never perfectly reached
(in finite times). It is therefore necessary to introduce a threshold to
decide whether the 0-state has been reached, with the related problem of
having to clarify whether the results are truly independent of
the threshold. In order to settle this issue, we find it convenient to determine
the Finite Amplitude Lyapunov Exponent (FALE) \cite{ABCPV96}. We do so
by first introducing $\tau(W)$, the average time needed by the field norm
\begin{equation}
||w(t)|| = \frac{1}{L} \sum_i^L |w_i|
\end{equation}
to become for the first time smaller than a preassigned threshold W. 

The FALE can be thereby defined with reference to
a sequence of exponentially spaced thresholds $W_n$ ($W_n/W_{n-1}=r<1$) as
\begin{equation}
\Lambda(W_n) =  \frac{\log r}{\tau(W_{n+1}-W_n)}  \, .
\end{equation}
In the limit $r\to 1$
\begin{equation}
\Lambda(W) =  \left[ \frac{d\tau(W)}{d\log W} \right]^{-1} \, .
\end{equation}
In the further limit $W\to 0$, $\Lambda(W)$ reduces to the usual Lyapunov
exponent.
\begin{figure}[ht]
  \begin{center}
    \includegraphics[clip=true,width=9cm]{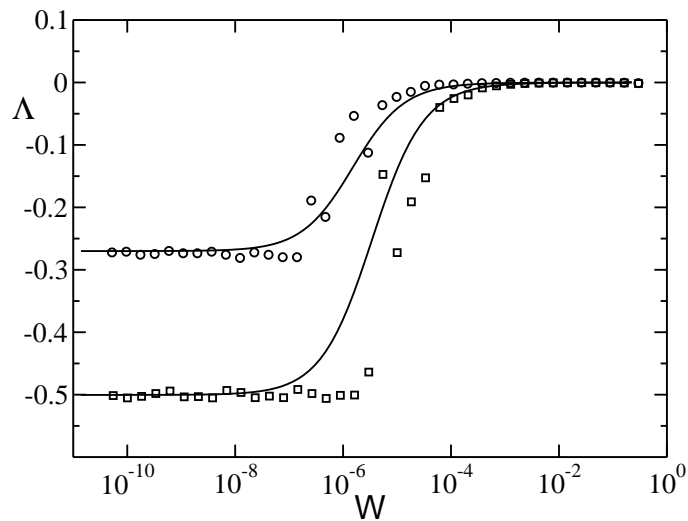}
  \end{center}
\caption{Finite amplitude  Lyapunov exponent of the stochastic model for two
different sets of parameter values, both at criticality: circles refer to
$\Delta=0.01$, $L=256$, $a_c=0.6055$, squares to $\Delta=0$, $L=128$, and
$a_c=0.6063$.}
\label{fig:fsle}
\end{figure}
In Fig.~\ref{fig:fsle}, we see that the FALE while being almost equal to zero at
sufficiently large scales, becomes equal to the true Lyapunov exponent
below a certain threshold $W_c$. Accordingly, since for $W<W_c$, the dynamics
is dominated by the usual Lyapunov exponent, we can safely conclude that when
the norm becomes smaller than $W_c$, the absorbing state will be reached with
probability one and this solves the problem of an unambigious identification of
the threshold. Moreover, detailed numerical simulations have revealed
that $W_c$ decreases faster than $1/L$, where $L$ is the system size \cite{GLPT03}.
In discrete-variable systems, the minimal non-zero value that $W$ can
meaningfully take is $1/L$ (which corresponds to just one active site). 
As $W_c<1/L$, one can conclude that in this stochastic system the
scaling range is even broader than in usual discrete systems.
Now a comment about the reason why the linear stability analysis may not apply
at vanishing distances. In fact, when the above defined norm of a vector is
small, the field can nevertheless be sporadically of ${\mathcal O}(1)$. 
The behaviour of such bursts may represent an obstruction to the validity of 
the linear stability analysis and this is what tells Fig.~\ref{fig:fsle}.

Finally, we recall that FALE have been employed to characterize single maps of
the type (\ref{eq:fun}) revealing that for sufficiently small $\eta$ and for some
finite $W$, the FALE is indeed larger (positive) than the standard Lyapunov
exponent \cite{cencini_torcini_pre}. Moreover, a generalization of the FALE to a
comoving reference frame allows to formulate a marginal stability criterion that
is able to predict the velocity on both cases of linear and nonlinear
propagation \cite{cencini_torcini_pre}. Moreover, coupled maps (\ref{eq:fun}) with
$\eta=0$ have been also analyzed by Letz \& Kantz \cite{Kantz} who introduced an
indicator similar to the FALE (i.e. able to quantify the growth rate of
non infinitesimal perturbations). This indicator turns out to be negative for
infinitesimal perturbations and becomes positive for finite perturbations.
This means that a sufficiently large perturbation can propagate along the
system due to nonlinear effects. This confirms previous observations
for marginally stable systems \cite{TGP95}.

\section{More realistic models}
\label{sec:mrm}

In order to test how general {\it stable chaos} is, it is natural to start by
asking when discontinuities or quasi-discontinuities can be expected to arise 
in the physical world. In fact, we have seen that the source of indeterminacy
is the sudden amplification of the distance between two nearby trajectories,
once they fall on opposite sides of a discontinuity. In such a case, no
matter how small the initial distance is, the separation is suddenly amplified
to a value of ${\mathcal O}(1)$, that is determined by the size of the
discontinuity.

\begin{figure}[ht]
  \begin{center}
    \includegraphics[clip=true,width=9cm]{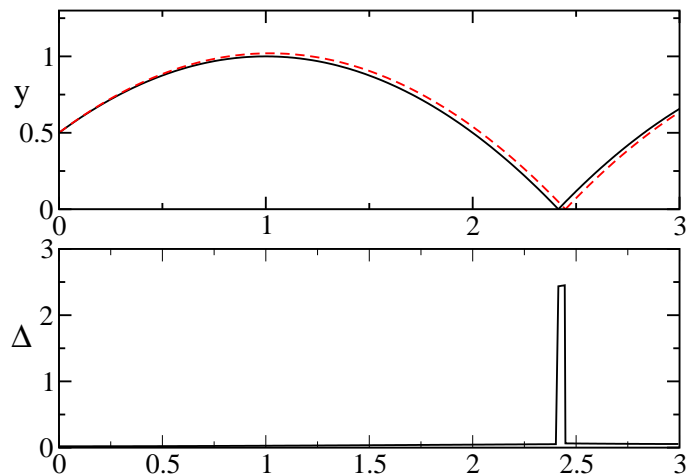}
  \end{center}
\caption{Separation between two nearly equal trajectories of a point particle 
colliding elastically against the floor. The distance
$\delta = sqrt(\delta y^2 + \delta v^2)$ is plotted versus time.
}
\label{fig:coll1}
\end{figure}

Before exploring the possible occurence of such phenomena, it is important to
stress that the discontinuity we are referring to is not a discontinuity in
time of the type associated, e.g., to collisions. A $\delta$-like collision
induces an abrupt change of a variable (specifically, the velocity), but this
affects the difference between two nearly identical trajectories only for a
short (infinitesinal) time lapse, after which the trajectories come close again.
This is illustrated in Fig.~\ref{fig:coll1}, where we have plotted the
time evolution of a point-like particle bouncing elastically on the
floor. In the lower panel the time evolution of the Euclidean distance
is represented: only within the short time window between the collisions of the
two trajectories with the floor, the relative distance becomes of order
${\mathcal O}(1)$. This is at variance with the map lattice model
(\ref{eq:map1},\ref{eq:fun}), where the distance, once amplified, remains large.

In the following two subsections, we illustrate some arguments
supporting the idea that a natural source of such a type of discontinuities is
associated with an exchange between non-commuting $\delta$-like events. 

\subsection{A Hamiltonian model: diatomic hard-point chain}

Before introducing the model, it should be remarked that in Hamiltonian
systems, conservation of volumes implies that the maximum Lyapunov exponent
cannot be negative; at most, all Lyapunov exponents are exactly equal to zero.
In fact, the Hamiltonian version of SC is the world of marginally
stable and yet ergodic models and it often goes under the name of {\it
pseudochaos} (see, e.g. \cite{falcioni}). Here, we are moslty interested in
emphasizing the analogies with SC and for this reason the diatomic
hard point gas (HPG) turns out to be rather appropriate also for its
relationship with billiard-like models, that are often invoked in the analysis
of pseudochaos.

\begin{figure}[ht]
  \begin{center}
    \includegraphics[clip=true,width=9cm]{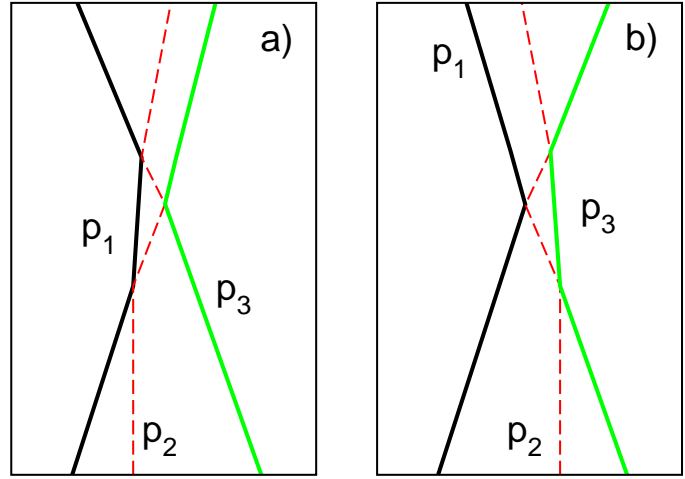}
  \end{center}
\caption{Evolution of two nearly equal set of initial condition in a
diatomic hard-point gas. By slihgly shifting the position of the middle
particle, the final velocities change abruptly when passing across the
three-body collision.
}
\label{fig:scatt}
\end{figure}
The diatomic HPG is a unidimensional system of point-like particles with
masses $m$ and $M$ that alternate along a line and undergo elastic collisions.
In the limit $m=M$, the model is perfectly integrable, since the velocities of
the two particles involved in any collision are simply interchanged.
Therefore there is no mechanism leading to a diffusion in velocity space.
However, as soon as the masses are assumed to be different, all numerical
simulations suggest that the dynamics is ergodic. On the other hand, it is easy
to convince oneself that the maximum Lyapunov exponent is still exactly equal
to zero. The argument is pretty straightforward \cite{grass02}: since the
collision rule is linear,
\begin{eqnarray}
u' &=&\frac{(m-M)u+2Mv}{m+M} \\
v' &=& \frac{2mu-(m-M)v}{m+M}
\end{eqnarray}
both real- and tangent-space dynamics follow the same rule. As a result,
the kinetic energy conservation ($\sum_i m_i v_i^2=E$, where $m_i = m$ or $M$,
depending on the parity of $i$) translates into the conservation of a suitably
weighted Euclidean norm of the perturbation field, namely,
$\sum_i m_i \delta v_i^2$. This means that the Euclidean norm of any vector is
conserved, irrespective of its direction, so that all Lyapunov exponents are
equal to zero. 

In the absence of deterministic chaos, which is, therefore, the source of the
stochastic-like behaviour exhibited by diatomic HPG chains? As illustrated in
Fig.~\ref{fig:scatt}, we argue that the source are the discontinuities occurring
around, e.g., three-body collisions. Let us consider an initial condition like
that in the left panel of Fig.~\ref{fig:scatt}: it gives rise to a sequence of
three collisions, $1-2$, $2-3$, $1-2$ before the particles separate out. By
shifting the position of the central particle (this is equivalent to moving the
initial $x_i$ variable in the CML), we pass to the condition depicted in the
right panel, which gives rise to the collisions $2-3$, $1-2$, $2-3$. Accordingly,
the sequence of two-body collisions changes abruptly in correspondence of a
three-body collision, when the three particles find themselves in the same
place at the same time. As a consequence of this sudden modification, the
three final velocities differ in the two cases, as it can be appreciated by
comparing the two panels in Fig.~\ref{fig:scatt}.  Only in the limit case of
equal velocities, there is no discontinuity, since the final set is the same
for both sequences. In the former case, when starting from the sequence
$v_1$, $v_2$, $v_3$, one passes first to $v_2$, $v_1$, $v_3$, then to
$v_2$, $v_3$, $v_1$, and finally to  $v_3$, $v_2$, $v_1$. One can easily verify
that the final state is the same in the latter case too.

\begin{figure}[t]
  \begin{center}
    \includegraphics[clip=true,width=9cm]{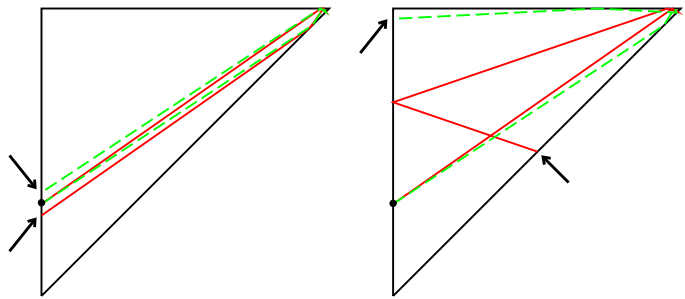}
  \end{center}
\caption{Billiard-like representation of the dynamics of a chain of two
particles moving in an interval with reflecting boundaries. Horizontal and
vertical axes correspond to the coordinates of the first and second particle
respectively. Two slightly different trajectories are plotted in each panel
until the 5th collision. Left and right panels correspond to $m=M$, and $M=2m$,
respectively.}
\label{fig:billi}
\end{figure}

The model dynamics can be further clarified by exploring the analogy with
billiard models. The connection was first discovered in Ref.~\cite{glashow},
where the authors considered the slightly different model of hard rods. Here, we
illustrate the relatively simple case of a gas of two particles $P_1$ and $P_2$
bounded to move between two fixed barriers $B_l$ and $B_r$, located in $x=0$ and
$x=1$, respectively. The linear position of the two particles can be represented
as the position of a point-particle in the plane and the constraints
$0\le x_1\le x_2\le 1$ imply that the motion is restricted to the triangular
region depicted in Fig.~\ref{fig:billi}. Collisions with the two mutually
orthogonal triangular edges correspond to collisions with either the left or
the right barrier, while those with the diagonal correspond to interparticle
collisions. Finally, the three angular points correspond to the only two
possible three-body collisions, $B_l P_1 P_2$, $P_1P_2B_r$ and to the synchronous
occurrence of the two-body collisions $B_lP_1$ and$P_2B_r$. In the equal mass
case, there is a perfect correspondence between hard point gas and the
triangular billiard. Accordingly, we can invoke the conjecture raised in
\cite{casati} that billiards with rational angles (expressed in $\pi$ units)
are necessarily ergodic. The crucial difference that appears as soon as the
two masses are assumed to differ from one another is that in the billiard-like
representation, the mass itself assumes a vectorial character. In particular,
incoming and outgoing velocities are not mutually symmetric in correspondence
of a collsion with the diagonal. However, the most relevant consequence is the
appearance of true discontinuities. This is illustrated by comparing two 
nearby trajectories which undergo a different sequence of collisions.
In the left panel of Fig.~\ref{fig:billi}, which refers to equal masses,
we see that the small difference in the initial velocity generates a slow
linear increase of the mutual distance. In the right panel, which refers to
$M=2m$, the two trajectories, although starting from the same initial
conditions, drastically separate out and find themselves very far apart 
after as few as 5 collisions (see the arrows).

\subsection{Neural networks}
A perhaps more interesting example of a model exhibiting stable chaos is
a network of leaky-integrate-and-fire neurons, where exponentially long
transients have been identified in various set-ups \cite{zumdieck,zillmer1,timme,memmer}. 
By following \cite{zillmer1}, the model dynamics for a network of
$N$ neurons can be written as a set of $N$ differential equations
\begin{equation}
\dot v_i = c - v_i - (v_i+w) \sum_{j=1}^N \sum_m g_{ij} \delta(t-t_j^{(m)})
\label{eq:lif}
\end{equation}
where the connectivity matrix $g_{ij}$ is defined as
\begin{equation}
\label{eq:connect}
g_{ij} = \begin{cases} 
G/\ell_i  &  \hbox{if $i$ and $j$ are coupled} \\ 
0  & \hbox{otherwise} ,
\end{cases}
\end{equation}
$\ell_i$ is the number of neurons that are connected to the $i$th neurons, and
$G$ is the coupling strength. Here, we will limit to consider the case of 
inhibitory coupling, that, in these notations, corresponds to a positive $G$
value. All variables are dimensionless and suitably rescaled: the
``action potential" $v_i \in [-\infty,1]$ whenever reaches
the limit value $v_j=1$, is reset to 0 and a $\delta$-spike is thereby
emitted and received by all the connected neurons. The parameter $c$ controls
the relaxation velocity, while $w$ quantifies indirectly the dependence of the
effect of the spike on the instantaneous value of the action potential.

\begin{figure}[ht]
  \begin{center}
    \includegraphics[clip=true,width=9cm]{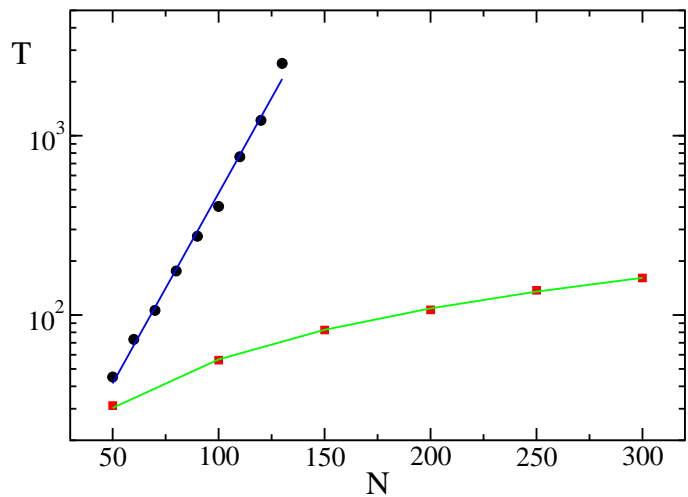}
  \end{center}
\caption{Transient length for the neural network model (\ref{eq:lif}) for $c=2$,
$w=4/7$ and 5\% of broken links. Squares (circles) correspond to $G=0.5$ and
($G=1.8$). The solid lines have been obtained by means of a linear and
exponential fit, respectively.}
\label{fig:trans}
\end{figure}

When all connections are active, the dynamics rapidly converges towards a
stationary state characterized by a sequence of evenly spaced spikes
(this is a so called splay state \cite{splay,zillmer2}). 
In the presence of disorder, such as, e.g.,
a small fraction of randomly broken links, the evolution may signficantly 
differ, depending on the coupling strength $G$. Below a certain critical
value, there is still a fast convergence towards an ordered state where the
neurons fire in a fixed order (in agreement with Jin's theorem
\cite{Jin:2002}); for sufficiently large coupling constants, the
average (over different realizations of the disorder) transient length $T$
is exponentially large with the number of neurons \cite{zillmer1}.
This is illustrated in Fig.~\ref{fig:trans}, where we have plotted the average
transient  for different system sizes: squares and circles correspond to
$G=0.5$ and 1.8, respectively. The solid lines are the result of a linear and
an exponential fit, respectively. The exponential increase of the transient is
a clear indication of SC, since at the same time, the maximum Lyapunov exponent
(after removing the zero exponent corresponding to a shift along the
trajectory) is definitely negative (as shown in \cite{zillmer1}).

\begin{figure}[ht]
  \begin{center}
    \includegraphics[clip=true,width=9cm]{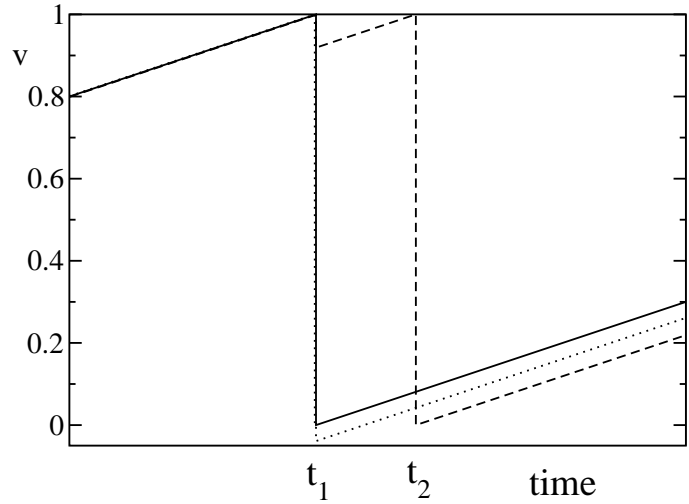}
  \end{center}
\caption{Evolution of two neurons with nearly the same potential and asymmetric
coupling: neuron $i$ couples to $j$, but not vice versa. The dotted
line corresponds to the evolution of neuron $i$; solid (dashed) curve
corresponds to the evolution of neuron $j$ in case its action potential is
initially smaller (larger) than that of neuron $j$.}
\label{fig:discn1}
\end{figure}

Which is the source of such long transients? In between the spikes, the single
potentials relax independently towards $c$  (a value that is not reached, since
$c>1$). Therefore, like in the diatomic hard point gas, the evolution is
piecewise linear and one can derive an analytic expression for the map as
rigorously done in Ref.~\cite{zillmer1}. In the absence of jumps between
different branches, the dynamics would be globally stable; the negativity of
the Lyapunov exponent is a reminiscence of such a stability. However, like in
the previous cases, there are discontinuities associated with abrupt changes in
the firing order of the neurons. Let us indeed consider two neurons $i$ and $j$
such that $g_{ji}=0$, while $g_{ij}\ne0$ and consider two different initial
conditions: (i)
$v_i(0)=v_j(0)-\e$, (ii) $v_i(0)=v_j(0)+\e$. A schematic view of the evolution is
presented in Fig.~\ref{fig:discn1}, where the solid line corresponds to the
dynamics of the $i$th-neuron, while dashed and dotted line denote the former
and latter trajectories, respectively. There, one can see that for times
larger than $t_2$ the two trajectories are separated by a finite distance,
as a result of a discontinuity in the dynamical law. This is due to the 
dependence of the inhibitory effect of a spike on the actual value of $v$ 
(see the multiplicative factor $(u+w)$ in Eq.~(\ref{eq:lif})). Being the 
size of the discontinuity of the same order of the coupling strength
(${\mathcal O}(1/N)$), one might argue that this is negligible for $N$
large enough. This is not the case, because it has to be compared with
the changes induced by the smooth dynamics in between two consecutive
spikes that is of the same order. Moreover, the distance between the two 
trajectories is even amplified to ${\mathcal O}(1)$ in the time 
interval $[t_1,t_2]$. As $t_2-t_1$ is, by definition of the same order of 
the interspike interval, this effect too is, in principle, nonnegligible.

More recently, stable and yet irregular behaviour has been reported also in the
context of a slightly different neural network, where the spike are assumed to
be received with a finite delay $\tau$ and the spike effects are independent
of $v$ (see \cite{timme}). From the point of view of {\it discontinuities}, 
this latter property inhibits a persistent amplification of distances between 
nearby trajectories. Nevertheless, the finite-time amplification mechanism is 
still present and the very fact that long-stable transients have been observed 
is an indication that it lasts enough to yield ``avalanches" and  thereby to 
a self-sustained irregular behaviour. However, one should also notice that
``discontinuities" are a necessary but not sufficient condition for the
onset of SC.

\section{Conclusions}
\label{sec:conclusion}

In the present Review we have thoroughly discussed the phenomenon of
{\it stable chaos}, a type of irregular behaviour occurring in deterministic
systems that manifests itself as an exponentially (with the system size) long
and stationary transient. SC differs from usual chaos in that it is 
characterized by negative Lyapunov exponents, but still reatins some 
features that are reminiscent of deterministic chaos. In fact, by
smoothing out the discontinuities present in the most typical SC models, 
induces the multifractal spectrum of the maximum Lyapunov exponent to
extend to positive values. This, in turn, suggests that topological chaos 
(i.e. a strictly positive topological entropy) is a prerequisite for the 
observation of SC. However, we have shown that linear stability 
analysis does not to provide a convincing description of relevant 
properties such as the propagation of finite perturbations. In this respect, 
a promising indicator is represented by the finite amplitude Lyapunov 
exponent, although there are conceptual difficulties in extending this 
approach beyond the maximum exponent.

A further interesting question concerns the transition from ordered behaviour
to SC. A detailed numerical analysis of a coupled-map lattice reveals the
existence of a fuzzy region, where behaviour that is neither strictly ordered
nor clearly chaotic has been detected. Is this just a difficulty due to strong
finite size effects, or this phenomenon hides the presence of a genuinely
``complex" (uncomputable) evolution? In an almost globally-coupled neural
network, the transition appears to be a standard point-like phenomenon, whose
universality class is however still unclear. 

The most important question concerns the generality of SC. All models where SC
has been observed do possess strong localized nonlinearities that may reduce to
true discontinuities in phase space. The first models where SC has been
observed are somehow artificial systems with no direct relationship with the
physical world. However, the discussion of the diatomic hard point gas and of
the network of pulse coupled neurons, has contributed to clarify that
discontinuities may spontaneously emerge in models characterized by the
presence of non communting ``$\delta$-like" events (such as two-body collisions
or spike emissions). Moreover, since we have seen that SC survives a smoothing 
of the coupled-map model, we may also conjecture that the same holds true in 
these latter contexts, once we assume finite collision times or finite
pulse-widths.

\section{Acknowledgment}
We would like to acknowledge those who have collaborated with us on this
problem over the years: R. Bonaccini, F. Cecconi, M. Cencini, F. Ginelli, 
P. Grassberger, R. Kapral, S. Lepri, R. Livi, G.L. Oppo, and R. Zillmer. 
Moreover, we wish to thank M. Timme, M. Wolfrum, and S. Yanchuk for recent 
useful discussions. This work has been partly carried out with the support of the
EU project NEST-PATH-043309.

\end{document}